\documentclass[apj,numberedappendix]{emulateapj}
\journalinfo{Accepted to the Astrophysical Journal}

\usepackage{rotating}

\shorttitle{The Mass-Size Relation from Clouds to Cores. I. A New
  Probe of Cloud Structure}

\shortauthors{Kauffmann et al.}

\begin{document}

\title{The Mass-Size Relation from Clouds to Cores. I. A new Probe of
  Structure in Molecular Clouds}

\author{J.\ Kauffmann\altaffilmark{1,2,3}, T.\
  Pillai\altaffilmark{2,3}, R.\
  Shetty\altaffilmark{1,2,3}, P.\ C.\ Myers\altaffilmark{2} \&
  A.\ A.\ Goodman\altaffilmark{1,2}}


\altaffiltext{1}{Initiative in Innovative Computing (IIC), 60 Oxford
  Street, Cambridge, MA 02138, USA}
\altaffiltext{2}{Harvard-Smithsonian Center for Astrophysics, 60
  Garden Street, Cambridge, MA 02138, USA}
\altaffiltext{3}{present addresses: Jens Kauffmann, NPP Fellow, Jet
  Propulsion Laboratory, 4800 Oak Grove Drive, Pasadena, CA 91109,
  USA; Thushara Pillai, California Institute of Technology, MC 249-17,
  1200 East California Boulevard, Pasadena, CA 91125, USA; Rahul
  Shetty, Zentrum f\"ur Astronomie der Universit\"at Heidelberg,
  Institut f\"ur Theoretische Astrophysik, Albert-Ueberle-Str.\ 2,
  D-69120 Heidelberg, Germany}

\email{jens.kauffmann@jpl.nasa.gov}

\begin{abstract}
  We use a new contour-based map analysis technique to measure the
  mass and size of molecular cloud fragments continuously over a wide
  range of spatial scales ($0.05 \le r / {\rm pc} \le 10$), i.e., from
  the scale of dense cores to those of entire clouds. The present
  paper presents the method via a detailed exploration of the Perseus
  Molecular Cloud. Dust extinction and emission data are combined to
  yield reliable scale-dependent measurements of mass.

  This scale-independent analysis approach is useful for several
  reasons. First, it provides a more comprehensive characterization of
  a map (i.e., not biased towards a particular spatial scale). Such a
  lack of bias is extremely useful for the joint analysis of many data
  sets taken with different spatial resolution. This includes
  comparisons between different cloud complexes. Second, the
  multi-scale mass-size data constitutes a unique resource to derive
  slopes of mass-size laws (via power-law fits). Such slopes provide
  singular constraints on large-scale density gradients in clouds.
\end{abstract}


\maketitle

\section{Introduction}
Some of the most fundamental properties of molecular clouds are the
mass and size of these clouds and their substructure. Today, these
properties are well constrained: we know the masses and sizes of dense
cores in molecular clouds ($\lesssim 0.1 ~ \rm pc$ size; e.g.\
\citealt{motte1998:ophiuchus}, \citealt{johnstone2000:rho_ophiuchi},
\citealt{hatchell2005:perseus}, \citealt{enoch2007:cloud_comparison}),
and those of clumps (some $0.1 ~ \rm pc$) and clouds
($\gtrsim 10 ~ \rm pc$) containing the cores (e.g.,
\citealt{williams1994:clumpfind}, \citealt{cambresy1999:extinction},
\citealt{kirk2006:scuba-perseus}; see \citealt{williams2000:pp_iv} for
definitions of cores, clumps, and clouds). We do, however, not know
much about the \emph{relation} between the masses and sizes of cores,
clumps, and clouds: traditionally, every domain is characterized and
analyzed separately.  As a result, it is still not known how the core
densities (and thus star-formation properties) relate to the state
of the surrounding cloud.

In principle, the relation between the mass in cloud structure at
large and small spatial scales is described by mass-size relations.
\citet{larson1981:linewidth_size} presented one of the first studies
of such relations. He concluded (in his Eq.\ 5) that the mass
contained within the radius $r$ obeys a power-law,
\begin{equation}
m(r) = 460 \, M_{\sun} \, (r / {\rm pc})^{1.9} \, .
\label{eq:mass-size-larson}
\end{equation}
Most subsequent work refers to this relation as ``Larson's
3$^{\rm rd}$ law'', and replaces the original result with
$m(r) \propto r^2$ (e.g., \citealt{mckee2007:review}). This ``law of
constant column density'' (with respect to scale, $r$) is now
considered one of the fundamental properties of molecular cloud
structure (e.g., reviews by \citealt{ballesteros-paredes2007:ppv},
\citealt{mckee2007:review}, \citealt{bergin2007:dense-core-review}).
This relation has, however, never been re-examined comprehensively on
the basis of up-to-date data. It is, e.g., not clear whether recent
dust extinction and emission work is consistent with
$m(r) \propto r^2$.

Further, the limitations of available structure identification schemes
(such as CLUMPFIND; \citealt{williams1994:clumpfind}) forced past
studies to break cloud structure maps up into discrete fragments.
These fragments typically have a size slightly larger than the map
resolution. As a consequence, the cloud structure is only probed in a
narrow spatial domain; the largest spatial features in a given map
are, for example, usually not characterized. Today, approaches
permitting automatic examination of a continuous range of spatial
scales are available. \citet{rosolowsky2008:dendrograms}, in
particular, provide software for such studies (their dendrogram
analysis); our work would be impossible without the work by
\citeauthor{rosolowsky2008:dendrograms}. Such software permits
derivation of spatially more comprehensive mass-size relations than
possible in the past.

In this series of papers, we combine contemporary column density
observations of high sensitivity with a new data analysis technique to
examine the mass-size relation in molecular clouds for a continuous
range of spatial scales of order $0.01 ~ {\rm to} ~ 10 ~ \rm pc$. We
rely on extinction maps of molecular clouds (here:
\citealt{ridge2006:complete-phase_i}), as well as maps of dust
emission \citep{enoch2006:perseus}. Following the terminology of
\citet{peretto2009:irdc-catalogue}, we define ``cloud fragments'' in
the maps as regions enclosed by a continuous column density contour,
and derive their mass and size at various contour levels. This is
implemented using algorithms introduced by
\citet{rosolowsky2008:dendrograms}.\medskip

The first two papers in this series establish our analysis approach
(part I, the present paper) and explore several clouds in the solar
neighborhood ($\lesssim 500 ~ \rm pc$; part II). Section
\ref{sec:method} of the present paper describes our cloud fragment
extraction and characterization scheme. In Sec.\
\ref{sec:mass-size-properties} we provide a first idea how basic
physical properties and observational limitations affect the mass-size
measurements. This includes a comparison to results obtained using the
CLUMPFIND algorithm. A detailed discussion of analysis uncertainties
is presented in Sec.\ \ref{sec:all-scales_perseus}. These are explored
using data for the Perseus Molecular Cloud. We also explain how dust
emission and extinction data can be combined for a given
cloud.

Section \ref{sec:purpose} briefly describes how the new map analysis
scheme might help to advance star formation research. As we describe
there, it will help to jointly analyse data taken at different spatial
resolution. This is a key feature in the age of multi-wavelength and
multi-resolution studies. We conclude with a summary in Sec.\
\ref{sec:summary}.

\begin{figure*}
\begin{tabular}{ll}
\large \textsf{a) input map} & \large\textsf{b) mass-size data}\vspace{-0.4cm}\\
\begin{minipage}{0.45\linewidth}
\includegraphics[width=0.7\linewidth,angle=-52.7]{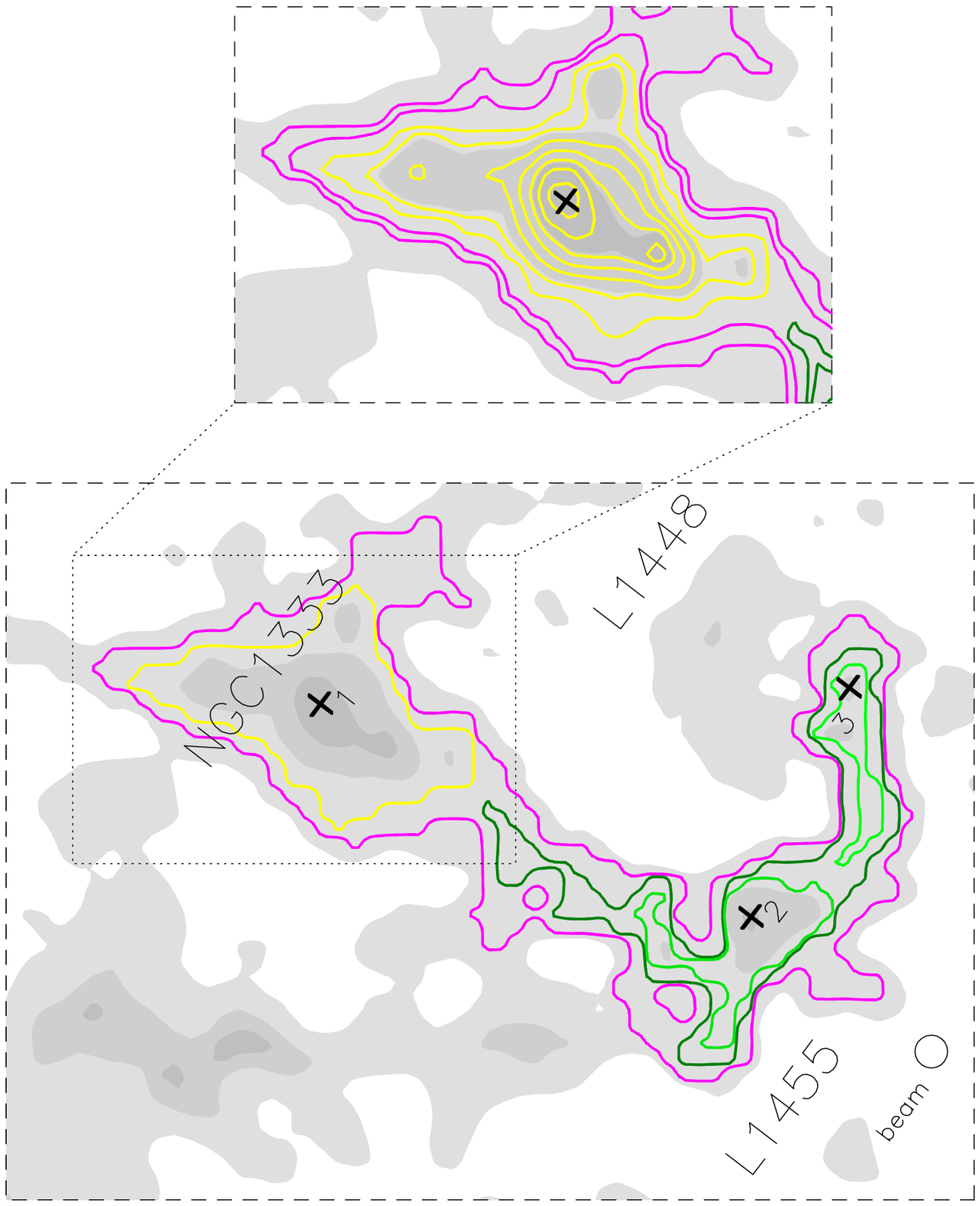}
\end{minipage} &
\begin{minipage}{0.45\linewidth}
\includegraphics[width=0.9\linewidth,bb=15 6 360 395,clip]{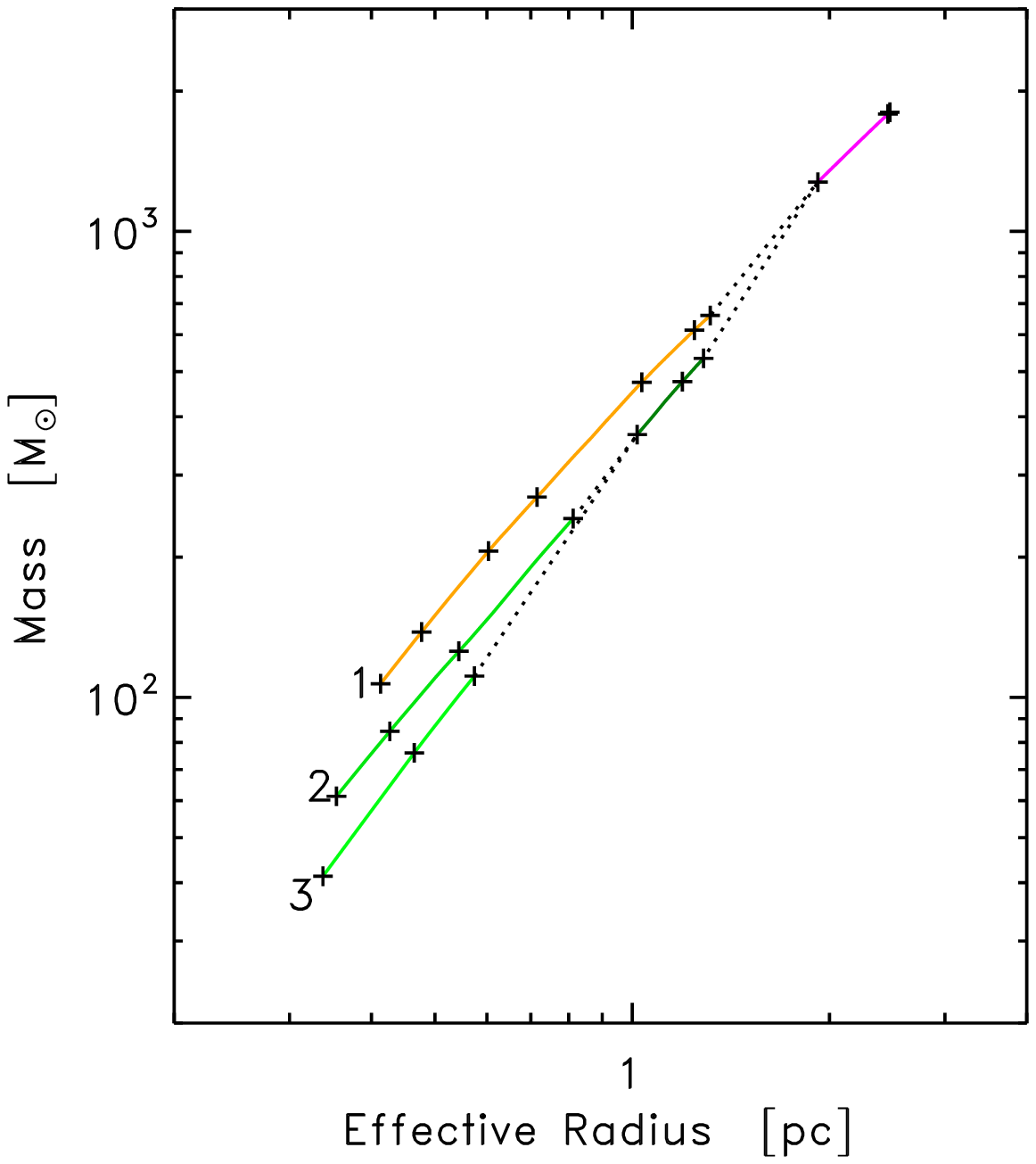}
\end{minipage}
\end{tabular}
\caption{Fundamental concept of mass-size measurements. Panel (a)
  shows an example column density map for the Perseus molecular cloud,
  where labels refer to individual star-forming regions. Such maps
  usually contain several local maxima (\emph{numbered crosses} in
  map). We pick one of these maxima, and draw contours at constant
  column density around this peak (\emph{map inset}). For each
  contour, we measure mass and size. These measurements can then be
  placed in a mass-size plot, as marked by \emph{crosses} in panel
  (b). When a map contains several maxima, it can be divided into
  cloud fragments where contours just contain a single maximum
  (\emph{light green} and \emph{yellow} boundaries in panel [a]), two
  of these (\emph{dark green} boundary), or even more (\emph{magenta}
  contour). In this sense, mass-size measurements for contours
  contained in one of these areas are related. These relations are
  highlighted by \emph{colored lines} in panel (b); the \emph{color}
  refers to the map region from where the measurements are taken, and
  \emph{numbers} indicate the central column density peak of each
  fragment. When two fragments blend into a combined one that contains
  both, mass and size measurements jump discontinuously from those of
  the individual fragments to those for the combined one. These jumps
  are indicated by \emph{dotted lines}.\label{fig:processing-scheme}}
\end{figure*}

\section{Method}
\label{sec:method}
\subsection{Processing of Contour Maps}
Consider some column density map containing a number of local maxima,
as sketched in Fig.\ \ref{fig:processing-scheme} (a). In this map,
cloud fragments can be identified as structures bound by some
continuous contour. To characterize these, we pick one of
the column density maxima, and measure the mass, $m$, and area, $A$,
contained within each column density contour containing this
peak. Here, we use the effective radius,
\begin{equation}
r = (A / \pi)^{1/2} \, ,
\end{equation}
to quantify $A$. By following the trends from contour to contour, as
shown in Fig.\ \ref{fig:processing-scheme} (b), one can construct a
mass-size relation for every cloud fragment. Strictly speaking, we do
thus construct mass-area diagrams. In practice, however, $r$ is
arguably a more intuitive variable than $A$.

Some contours may contain several local maxima. They represent
composite fragments. To give an example based on Fig.\
\ref{fig:processing-scheme}, the region bound by the dark-green
contour consists of the two regions marked in lighter shades of green,
and the magenta contour contains the fragments marked in yellow and
dark-green. Merging of two fragments occurs at the first column
density contour containing both of the fragments. In our analysis, we
enforce that only two fragments can merge at a time.

During a merger, mass and size jump discontinuously from the
pre-merger situation, $m_i$ and $r_i$, to their post-merger value,
$m = m_{\rm A}+ m_{\rm B}$ and $r = (r_{\rm A}^2 + r_{\rm B}^2)^{1/2}$. This
yields gaps in the mass-size relation (Fig.\
\ref{fig:processing-scheme} [b]). The merger information is preserved
during processing, so that it is possible to look up which fragments
are contained within others.

Our map characterization scheme is thus closely related to the one
employed by \citet{peretto2009:irdc-catalogue}. Like us, these authors
measure sizes and masses for regions bound by lines of constant column
density. The schemes differ in the number of contours considered: we
use a very large number ($10^2$ to $10^3$), where
\citeauthor{peretto2009:irdc-catalogue} only consider two contours per
cloud.\medskip

In practice, we use the dendrogram (i.e., tree analysis) code
presented by \citet{rosolowsky2008:dendrograms} for automatic
processing of the maps. A minimum significant contour has to be set
for every region; emission below this limit is not characterized here.
It is further necessary to specify a minimum contrast between peaks,
in order to identify significant central maxima for the objects. These
threshold column densities and contrasts are listed separately for
every region in the following. The identified maxima are required to
be spaced by more than one spatial resolution element. Again, this
parameter is noted separately for every map. In this work, all sources
with an effective diameter (i.e., $2 r$) smaller than twice the map
resolution are rejected and are treated as parts of enveloping
objects.

In the terminology of \citeauthor{rosolowsky2008:dendrograms}, our
mass measurement approach (i.e., integrate column density within some
contour) corresponds to their ``bijection paradigm''. Such mass
measurements, e.g.\ made towards a dense core, always give a sum over
several spatially overlapping components (e.g., some fraction of the
dense core, plus some fraction of its envelope). This is not a
problem, if properly taken into account in the later analysis. Other
choices are possible (e.g., the ``clipping paradigm'' of
\citeauthor{rosolowsky2008:dendrograms}), but they are less intuitive
and even harder to model. Also, most previously existing data has been
published effectively adopting the bijection paradigm.

\subsection{Mass Estimates\label{sec:mass-estimates}}
As a first example, below we present a mass-size analysis of Perseus,
based on column densities derived from 2MASS-derived extinction
data. As explained by \citet{ridge2006:complete-phase_i}, the map is
derived in terms of magnitudes of visual extinction, $A_V$. We convert
this to $\rm H_2$ column densities using the relation
\begin{equation}
N_{\rm H_2} = 9.4 \times 10^{20} ~ {\rm cm^{-2}} \, (A_V / {\rm
  mag})
\label{eq:coldens-dust-extinction}
\end{equation}
\citep{bohlin1978:av_conversion}. Mass surface densities, $\Sigma$,
can then be derived as
\begin{equation}
\Sigma = \mu_{\rm H_2} \, m_{\rm H} \, N_{\rm H_2} \, ,
\end{equation}
where $\mu_{\rm H_2} = 2.8$ is the mean molecular weight per $\rm H_2$
molecule \citep{kauffmann2008:mambo-spitzer} and $m_{\rm H}$ is the
weight of the hydrogen molecule. In practice,
$\Sigma = 0.047 ~ {\rm g \, cm^{-2}} \,
(N_{\rm H_2} / 10^{22} ~ {\rm cm^{-2}}) =
226 ~ M_{\sun} \, {\rm pc^{-2}} \,
(N_{\rm H_2} / 10^{22} ~ {\rm cm^{-2}})$, or
$A_V = 227.5 ~ {\rm mag} \, (\Sigma / {\rm g \, cm^{-2}})$. The mass
is then derived by integrating the mass surface density,
$m = \int \Sigma \, {\rm d} A$. For this we adopt a Perseus distance
of $260 ~ \rm pc$ \citep{cernis1993:distance-perseus}.

\citet{goodman2008:column-density} present a comparison of column
density tracers for the Perseus region. After studying column density
maps based on dust extinction (from 2MASS data), dust emission (from
IRAS imaging), and line emission data (from large field $\rm ^{13}CO$
[1--0] mapping), they conclude that extinction-based maps provide the
best available information on a cloud's spatial mass
distribution. Extinction-based column densities deviate by $\sim 25\%$
from those derived from other tracers (after removing global offsets
between estimates, e.g.\ due to the choice of dust opacities). The
true column density is supposedly in between these estimates. If every
tracer has a similar scatter with respect to the true column density,
this scatter would then be $\sim 25\% / 2^{1/2} \approx 18\%$ for all
tracers. Extinction-based estimates of the column density do probably
deviate by a lower amount from the true value. Here, we thus adopt a
systematic uncertainty of $\lesssim 15\%$ for extinction-based mass
estimates in Perseus.

Young stars embedded in the clouds can further bias extinction
observations, given their red intrinsic colors. This bias is
particularly significant towards clusters, such as NGC1333 and IC348
in Perseus. We do not exclude these regions from our study, but
measurements towards the clusters should be interpreted with
particular caution.

\subsection{Example Map\label{sec:example-map}}
\begin{figure}
\includegraphics[height=\linewidth,angle=-90]{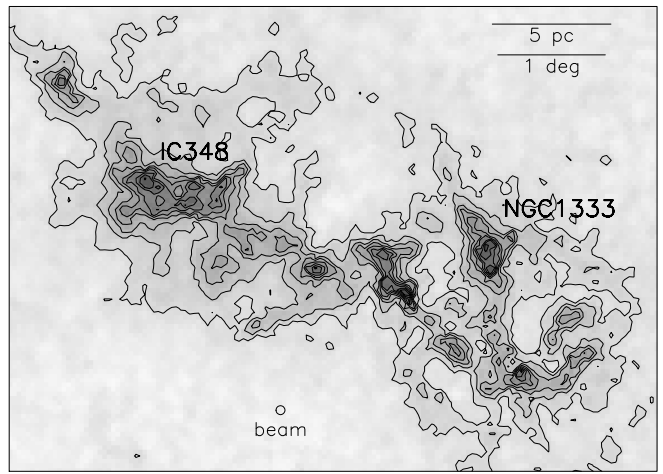}
\caption{Example column density map for Perseus, presented in terms of
  visual extinction, $A_V$. The map is taken from
  \citet{ridge2006:complete-phase_i}. \emph{Contours} are drawn in steps of
  $1 ~ \rm mag$, starting at $2 ~ \rm mag$. \emph{Labels} indicate the
  rough position of the star forming regions NGC1333 and
  IC348.\label{fig:example_map}}
\end{figure}

Figure \ref{fig:example_map} shows the aforementioned extinction map
for Perseus. The two most prominent star-forming regions in Perseus,
NGC1333 and IC348, manifest as extended extinction structures in this
map. Fragments of very small size (e.g., $\lesssim 0.1 ~ \rm pc$), like
dense cores around individual young stellar objects, are not visible in
the map, due to its too poor spatial resolution ($5 \arcmin$,
corresponding to $\approx 0.4 ~ \rm pc$). Section
\ref{sec:combine-masses} shows how such structures can still be
bootstrapped into mass-size studies.

\begin{figure}
\includegraphics[width=\linewidth,bb=20 54 361 386,clip]{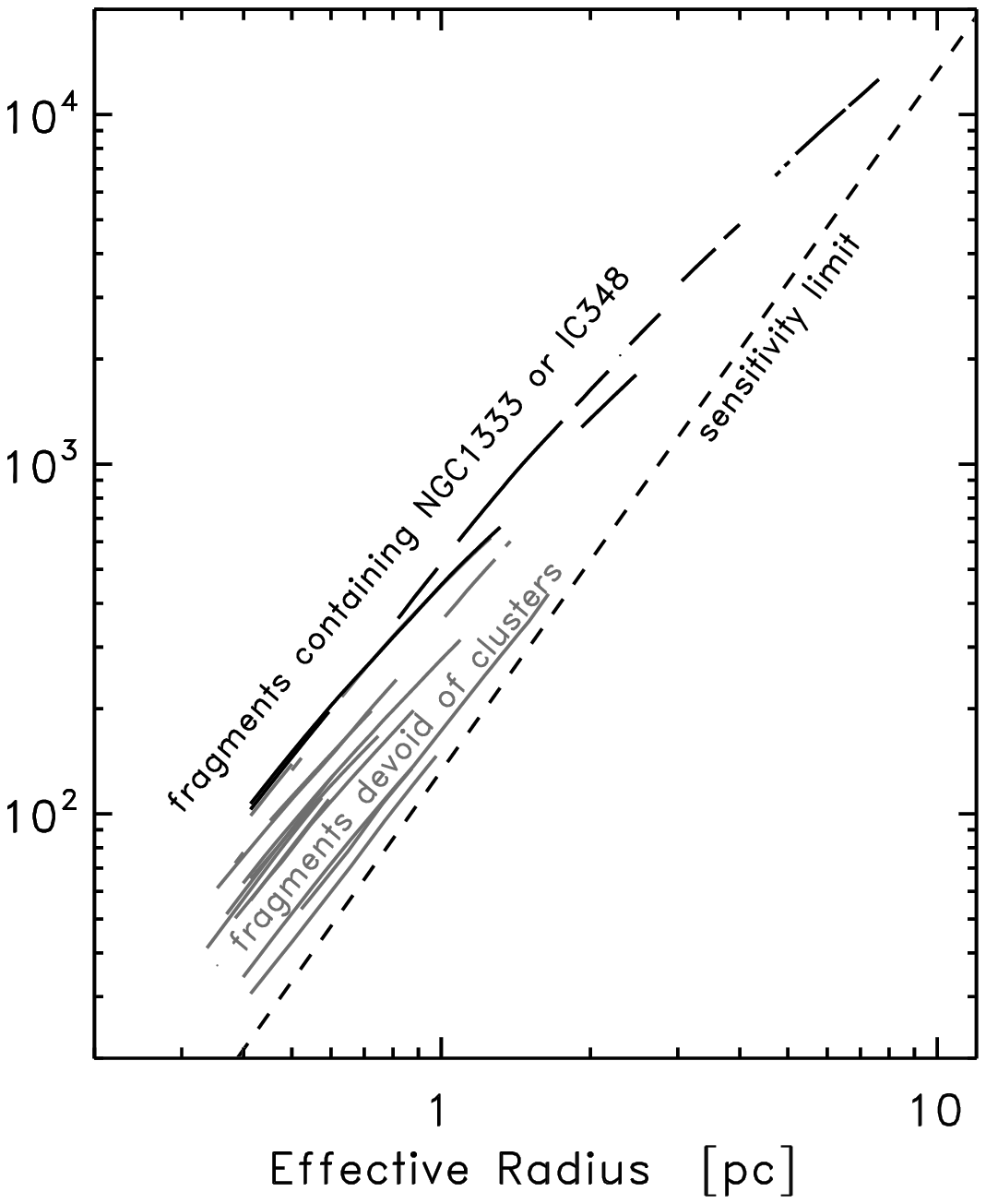}\\
\includegraphics[width=\linewidth,bb=20 10 361 386,clip]{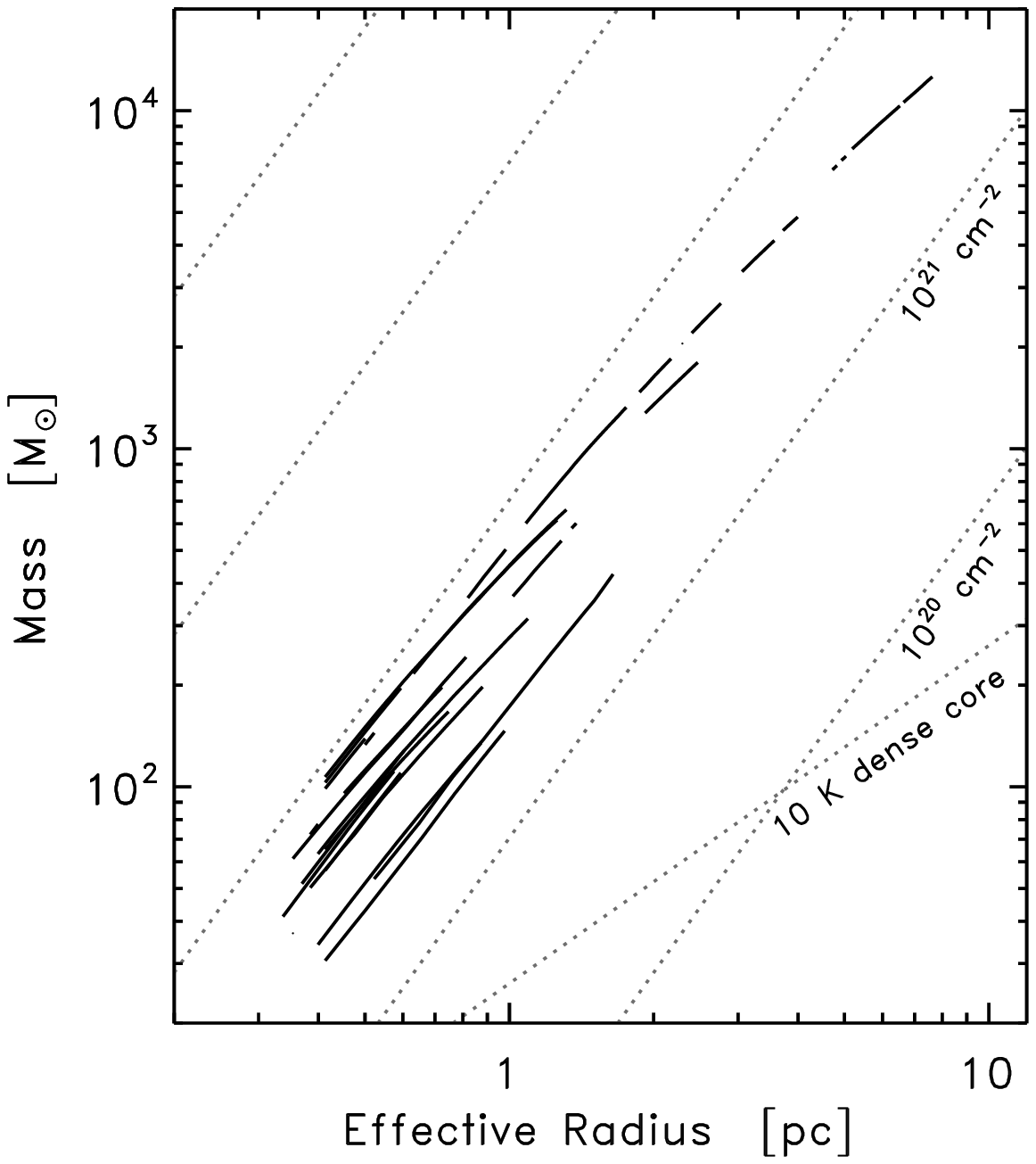}
\caption{Mass-size relation for the Perseus molecular cloud. Compared
  to Fig.\ \ref{fig:processing-scheme}, we only keep the lines
  connecting related measurements. The \emph{top panel} describes the
  data in detail. \emph{Bold solid lines} indicate data for fragments
  either containing NGC1333 or IC348. \emph{Light solid lines} show
  observations for regions not associated with these clusters. The
  \emph{dashed line} gives the sensitivity limit of the analysis. In
  the \emph{bottom panel} we present the data in the context of
  reference mass-size relations. \emph{Dotted lines} indicate mean
  $\rm H_2$ column densities, $\langle N_{\rm H_2} \rangle$,
  respectively the mass-size relation for a singular hydrostatic
  equilibrium sphere supported by isothermal pressure from gas at
  $10 ~ \rm K$ temperature (see Sec.\ \ref{sec:reference-relations}
  for both).\label{fig:example_mr-relation}}
\end{figure}

Figure \ref{fig:example_mr-relation} presents the mass-size data
derived from the Perseus extinction map. This diagram reveals that the
Perseus complex has a total (effective) radius of $8 ~ \rm pc$ and a
total mass of $1 \times 10^4 ~ \rm M_{\odot}$. The properties from
contours containing the two major stellar clusters, NGC1333 and IC348,
are highlighted by bold black lines. As one may naively expect, the
cloud fragments enclosing these clusters are, at given radius, the
most massive fragments within the cloud complex. This analysis also
reveals fragments that are, again at given radius, much less massive
than the regions containing the clusters. As discussed in the next
paragraphs, the column density sensitivity of the map sets a
radius-dependent lower limit to the masses that can be detected in a
given map. Fragments of a mass much lower than those shown here may
thus well exist in Perseus.  Eventually, at sufficiently large radius,
most of the contour-bound objects found merge into a single fragment
containing essentially all of the Perseus molecular cloud.

In the processing of the map, we have used a minimum threshold
extinction of $2 ~ \rm mag$. This limit is much larger than the noise
level of $0.4 ~ \rm mag$ \citep{ridge2006:complete-phase_i} and
rejects the signal from background extinction structures not related
to the cloud. Correspondingly, we cannot detect fragments with a mean
extinction below $2 ~ \rm mag$. Given the relation between mass, size
and mean column density (Eq.\ \ref{eq:mr_column-density}), this minimum
column density sets a lower limit to the detectable mass. Substitution
of the minimum column density, $N_{\rm H_2, min}$ (in Eq.\
\ref{eq:mr_column-density}; note that
$\langle N_{\rm H_2} \rangle \ge N_{\rm H_2, min}$), gives sensitivity
limits like the one shown in Fig.\ \ref{fig:example_mr-relation}.
The minimum contrast between peaks is required to be the noise level
times a factor 3, i.e.\ $1.2 ~ \rm mag$. The spatial resolution of the
map is $5 \arcmin$; we do therefore require the maxima to be separated
by at least 3 pixels ($7 \farcm 5$), and objects with a radius smaller
than $5 \arcmin$ are rejected as being unphysical.

\section{Properties of Mass-Size Data}
Some mathematical and physical laws governing the properties of
mass-size data have to be heeded, if a meaningful interpretation of
the observations is desired. Here we list the most fundamental of
these.\label{sec:mass-size-properties}

\subsection{Reference Relations\label{sec:reference-relations}}
A number of reference mass-size relations can help to navigate within
the observational data more intuitively. They are in part derived
assuming a spherical geometry for cloud fragments. The assumption of
spherical symmetry may not be appropriate, though. This caveat should
be kept in mind when using the following reference relations.

Mass and size measurements can be used to calculate the mean mass
surface (or column) density of a fragment,
$\langle \Sigma \rangle = m(r) / A(r)$. Conversely, one can draw lines
of constant mass surface density,
$m(r) = \langle \Sigma \rangle \, \pi \, r^2$. Conversion to column
density, and substitution of the aforementioned constants, yields
\begin{equation}
m(r) = 71 \, M_{\sun} \,
(\langle N_{\rm H_2} \rangle / [10^{21} ~ {\rm cm^{-2}}]) \,
(r / {\rm pc})^2 \, .
\label{eq:mr_column-density}
\end{equation}
These lines are drawn in most mass-size plots presented here (e.g.,
Fig.\ \ref{fig:example_mr-relation}). Equation
(\ref{eq:mr_column-density}) implies one of the most important
properties of mass-size data: since column density decreases with
increasing radius\footnote{By definition, this is always the case for
  our source characterization scheme. We start off from a column
  density peak and then consider, by design, regions that increase
  with size when lowering the threshold column density.}, the observed
mass-size relations must be flatter than $m(r) \propto r^2$.

Equation (\ref{eq:mr_column-density}) can actually be used to
transform our results into the analytical diagram presented by
\citeauthor{tan2007:massive-sf} (\citeyear{tan2007:massive-sf};
column-density vs.\ mass). Our study goes beyond the work by
\citet{tan2007:massive-sf} in that it systematically populates the
parameter space with observational data.

For spherical clouds, the mean density is
$\langle \varrho \rangle = m(r) / (4/3 \, \pi \, r^3)$. The
corresponding mass-size relation reads
$m(r) = 4/3 \, \pi \, \langle \varrho \rangle \, r^3$, or
\begin{equation}
m(r) = 282 \, M_{\sun} \, (\langle n_{\rm H_2} \rangle / [100 ~ {\rm cm^{-3}}]) \,
(r / {\rm pc})^3 \, ,
\end{equation}
where we substitute the density of $\rm H_2$ molecules,
$n_{\rm H_2} = \varrho / (\mu_{\rm H_2} \, m_{\rm H})$.

In part II of this series, we shall study spherical power-law
density profiles, $\varrho(s) \propto s^{-k}$ (where $s$ is the
radius), as models for the observed mass-size relations. We show that
\begin{equation}
\varrho(s) \propto s^{-k} \quad \Leftrightarrow \quad
m(r) \propto r^{3-k} \, .
\label{eq:slope-vs-density}
\end{equation}
The slope of the mass-size relation is therefore related to the slope
of the density law. A density profile $\varrho(s) \propto s^{-2}$ is
often adopted to describe dense cores (see
\citealt{dapp2009:density-profile} for a discussion). This gives a
mass-size relation $m(r) \propto r$. In hydrostatic spheres supported
by isothermal pressure, mass, size, and gas temperature are related by
\begin{equation}
m(r) = 2.6 \, M_{\sun} \,
\left( \frac{T_{\rm g}}{10 ~ \rm K} \right) \,
\left( \frac{r}{0.1 ~ \rm pc} \right)
\label{eq:mass-size-sis}
\end{equation}
(see \citealt{kauffmann2008:mambo-spitzer}, Eq.\ 13). For gas
temperatures $T_{\rm g} = 10 ~ \rm K$, one obtains the model relation
drawn in most mass-size diagrams of this paper (e.g., Fig.\
\ref{fig:example_mr-relation}).

We stress that we obtain two-dimensional mass-size relations from
column density maps. These are related to, but not identical with,
mass-size laws obtained from three-dimensional density maps. This is
illustrated by the experiments conducted by
\citeauthor{shetty2009:ppp-ppv} (\citeyear{shetty2009:ppp-ppv},
submitted) who use the fragment identification technique also used by
us. Their analysis is based on three-dimensional numerical simulations
of turbulent clouds. As part of their experiments, they fit power-laws
(similar to Eq.\ \ref{eq:mass-size-larson}) to their mass-size
data. For their particular set of simulations, the exponent derived in
this fashion is similar to the number of dimensions used for mass
measurements (i.e., 3 when based on density, and 2 when based on
column density). This underlines that the number of dimensions
considered has to be kept in mind. Note, though, that these details do
not compromise mass-size measurements as a tool for cloud structure
analysis. \emph{Observed} mass-size laws unambiguously summarize
actual cloud structure. Only their \emph{interpretation} is sensitive
to the assumed geometry.

\subsection{Relation to CLUMPFIND-like Results}
The approach chosen here constitutes one of several possible choices
to measure the mass and size of objects in maps. Another popular
approach is to use the CLUMPFIND algorithm
\citep{williams1994:clumpfind}. This method uses contours to break
emission up into several objects, just as done here.
CLUMPFIND-extracted boundaries do, however, not necessarily follow
contours. This is a major difference to our method, where objects are
always bound by some column density contour. Based on this fact alone,
CLUMPFIND and our approach will thus extract very different
objects. \citet{goodman2008:nature} illustrate this problem
comprehensively.

Further, CLUMPFIND does not allow for hierarchical structure, i.e.,
fragments containing other fragments. For example, CLUMPFIND will
never determine the integral properties of the entire cloud, since the
cloud is usually broken up into many independent fragments. The
relation of CLUMPFIND-results to cloud hierarchy is studied by
\citet{pineda2009:clumpfind}.

\begin{figure}
\includegraphics[width=\linewidth,bb=15 6 360 395,clip]{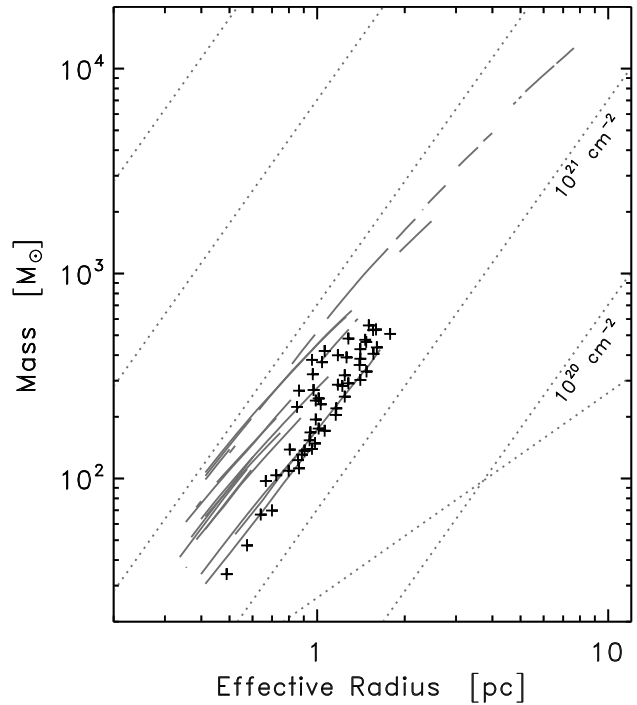}
\caption{CLUMPFIND (\emph{crosses}) and dendrogram mass-size results
  (\emph{lines}) for the extinction map presented in Fig.\
  \ref{fig:example_map}. The two approaches yield broadly similar,
  but not directly related results.\label{fig:comp-clumpfind}}
\end{figure}

Figure \ref{fig:comp-clumpfind} compares mass and size measurements
from CLUMPFIND to those from our approach\footnote{The CLUMPFIND
  results were kindly provided by J.E.\ Pineda.}. Both
characterizations are based on our Perseus extinction map. To initiate
CLUMPFIND, we choose contours as close as possible to the parameters
used in our contour-based segmentation (Sec.\ \ref{sec:example-map});
the contour spacing is $1.2 ~ \rm mag$, with the lowest contour at
$2 ~ \rm mag$. Still, this CLUMPFIND segmentation yields objects that
are only remotely related to the regions extracted by our method. To
start with, CLUMPFIND extracts 52 peaks enclosed by individual clumps,
where our method only identifies 17 significant peaks. This
discrepancy is a consequence of CLUMPFIND's relaxed peak
identification criterion: every local peak encircled by a continuous
contour is considered significant, independent of the depth of the
column density dip towards the next peak. There is thus no
well-defined correspondence between regions identified by the
different algorithms.

To characterize the relation between objects extracted by different
methods in some sense, we inspect the structure around the 17
significant column density peaks found by our method. For every such
peak, we identify the CLUMPFIND object containing this peak. The mass
of the latter clump can then be compared to our results, taken at the
clump's radius. Because of merging of objects, our method does not
provide a mass measurement for every possible radius (consider, e.g.,
the evolution of mass-size measurements starting from peak 3 in Fig.\
\ref{fig:processing-scheme}[b]). For those cases where mass
measurements exist for the CLUMPFIND-derived radius, we find that
CLUMPFIND gives masses of order 55\% to 95\% of the masses derived by
our approach.

In summary, mass and size measurements from CLUMPFIND are thus broadly
compatible with our results. By this we mean that the
CLUMPFIND-derived mass-size measurements reside in the space spanned
by our own measurements. There is, however, no good correspondence on
a detailed level.

\section{Perseus in Detail}
Given the tools derived above, it is now possible to characterize the
mass-size relation for Perseus. While doing so, we also evaluate
uncertainties affecting our analysis. In a final step, we extend the
analysis to clouds other than Perseus.\label{sec:all-scales_perseus}

\subsection{Mass Uncertainty\label{sec:uncertainty-mass}}
\begin{figure}
\begin{tabular}{lc}
\includegraphics[width=0.85\linewidth,bb=15 31 360 395,clip]{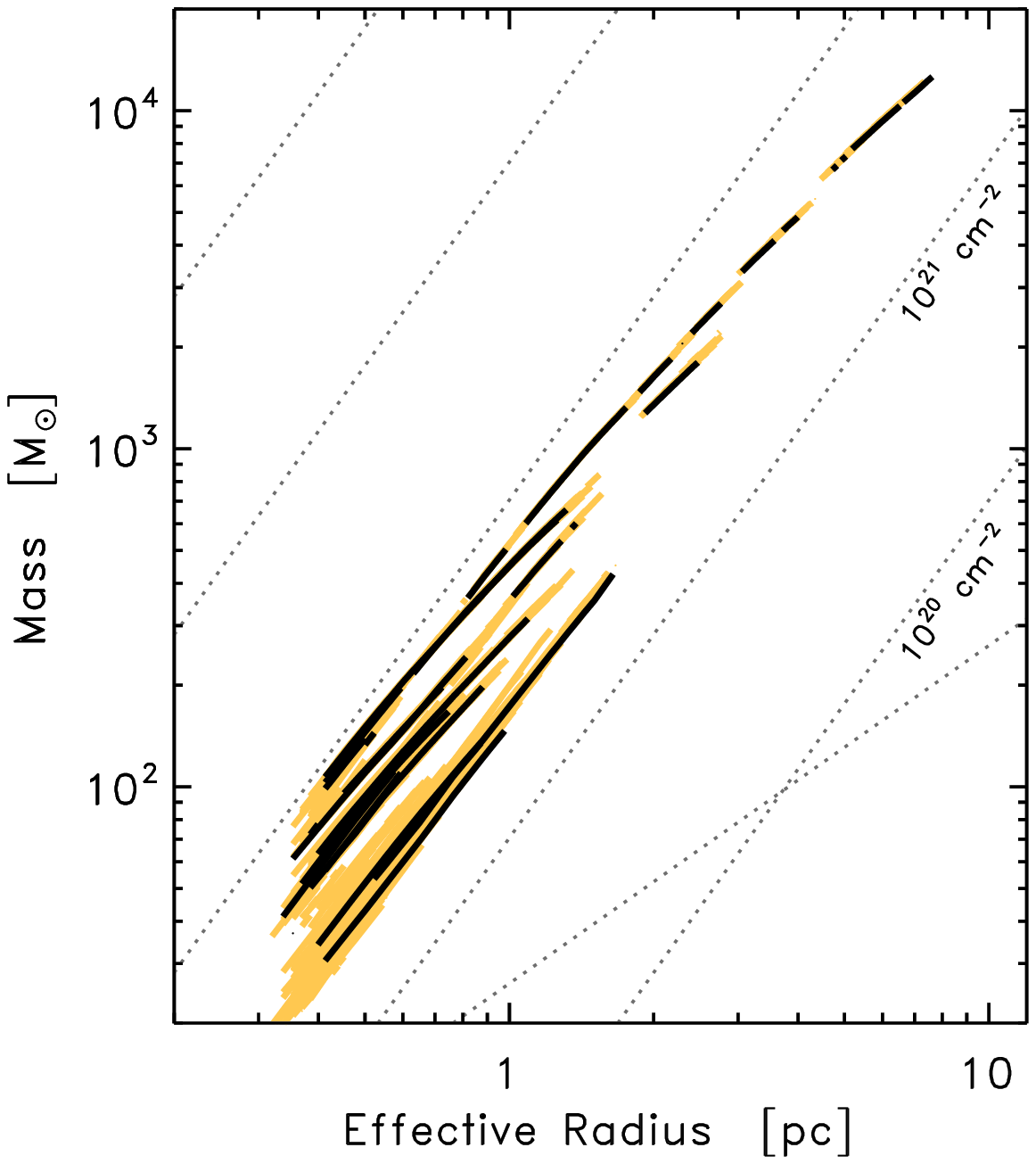} &
\begin{sideways}
\textsf{\hspace{1.8cm}\large{}(a) impact of noise}
\end{sideways}\\
\includegraphics[width=0.85\linewidth,bb=15 6 360 390,clip]{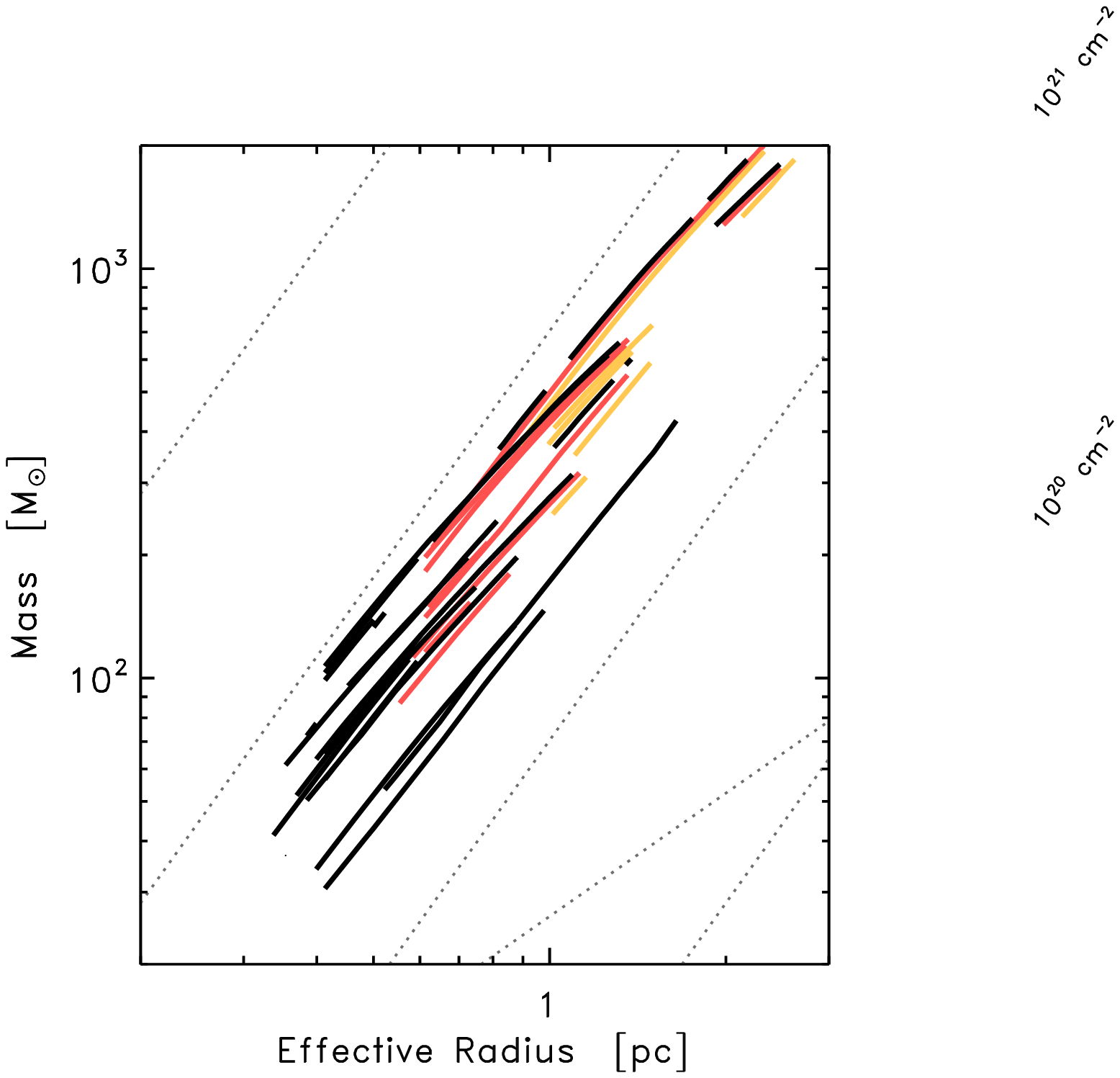} &
\begin{sideways}
\textsf{\hspace{2.5cm}\large{}(b) impact of smoothing}
\end{sideways}
\end{tabular}
\caption{Impact of noise (\emph{top}) and limited spatial resolution
  (\emph{bottom}) on mass-size measurements. \emph{Note that the size
    and mass scale differ between the panels.} Noise and resolution
  are explored using artificial maps created by adding noise to
  (\emph{top}), respectively smoothing of (\emph{bottom}), the
  observed data for Perseus (Sec.\ \ref{sec:uncertainty-mass}).
  Differences between the input data and the properties of the
  artificial maps are due to these biases. The original input data is
  drawn in \emph{black}.  \emph{yellow lines in the top panel}
  show mass-size relations from maps with additional noise similar
  to the observed one ($0.4 ~ \rm mag$). \emph{Red and yellow
    lines in the bottom panel} present mass-size relations for maps
  with a resolution worsened to
  $r_{\rm beam, f} / r_{\rm beam, i} = 3/2$ and $5/2$, respectively
  (and features with $r < 2 \, r_{\rm beam, f}$ removed). At given
  radius, both biases just induce moderate changes in mass (Sec.\
  \ref{sec:uncertainty-mass}).\label{fig:comparison-noise-resolution}}
\end{figure}

To explore the impact of noise on mass measurements, we run
Monte-Carlo experiments in which we add artificial Gaussian noise of
root-mean-square (RMS) amplitude $\sigma(m_i)$ to the map
before structure characterization (where $m_i$ is the mass per
pixel). Such trials are presented in Fig.\
\ref{fig:comparison-noise-resolution}(a). Comparison of the derived
mass-size data to the one from the original map does then reveal the
impact of noise. In our experiments, we find that
\begin{equation}
\frac{\sigma(m)}{m} = 6 \,
\frac{\sigma(m_i)}{m} \, \frac{r}{r_{\rm beam}}
\label{eq:uncertainty-noise}
\end{equation}
is an upper limit to the noise-induced mass changes, $\sigma(m)$. In
this, $r_{\rm beam}$ is the beam radius and the numerical constant
does slightly depend on the number of pixels per beam.

Equation (\ref{eq:uncertainty-noise}) follows from Gaussian error
propagation of $m = \Sigma_i m_i$. The numerical constant is increased
by a factor 3, though, to provide a strict upper limit to
uncertainties. To test for non-Gaussian sources of error, we validate
Eq.\ (\ref{eq:uncertainty-noise}) by comparing the original data with
those derived from maps with additional noise. For any given
structural branch with $r < 2 ~ \rm pc$, Eq.\
(\ref{eq:uncertainty-noise}) is indeed found to set an upper limit to
the mass deviation at given radius, following the noise experiments
depicted in Fig.\ \ref{fig:comparison-noise-resolution}(a). For larger
radii, a small (but significant) additional uncertainty of $\sim 1 \%$
has to be included to capture non-Gaussian sources of error.

In our study, the relative uncertainty of column density estimates
(i.e., $\sigma(m_i) / m_i$) is $\le 1/5$ in characterized regions
(given extinction noise levels and selection thresholds of
$0.4 ~ \rm mag$ and $2.0 ~ \rm mag$, respectively). Further, the mass
is at least as large as the one at the sensitivity limit (Fig.\
\ref{fig:example_mr-relation}[a]),
$m > \pi \, m_{i, \rm min} \, (r / r_{\rm beam})^2$ (in our map, the beam
contains, by area, $\pi$ pixels). Substitution of these parameters in
Eq.\ (\ref{eq:uncertainty-noise}) give uncertainties of 19\% for the
smallest extracted features (for which $r = 2 r_{\rm beam}$) located
just at the sensitivity limit. Since
$\sigma(m) / m \propto \langle N_{\rm H_2} \rangle^{-1} \, r^{-1}$
(via substitution of Eq.\ [\ref{eq:mr_column-density}]),
larger regions well above the detection threshold will suffer lesser
uncertainties, $\ll 10\%$.\medskip

In later papers, we will compare the properties of clouds that are
located at different distances. Since the angular resolution of the
observations is about constant, an increase in distance implies a
decrease in physical resolution. To first order, this decrease in
physical resolution corresponds to smoothing of the map. We explore
the impact of smoothing by extracting mass-size data from smoothed
maps of the Perseus cloud, as demonstrated in Fig.\
\ref{fig:comparison-noise-resolution}(b).

Smoothing does effectively mean that mass is transferred to larger
scales. For fixed radius, smoothing does thus imply a reduction in
mass. To estimate this reduction, consider a region of effective
radius $r$. After smoothing, the mass contained in this
region will be smeared our over an area of radius
$r' \sim r_{\rm ini} + r_{\rm k}$. In this,
$r_{\rm k} = (r_{\rm beam, f}^2 - r_{\rm beam, i}^2)^{1/2} /
(\ln(2))^{1/2}$ is the effective radius of the smoothing kernel
required to go from the initial to the final beam radius of the map
during smoothing, $r_{\rm beam, i} \to r_{\rm beam, f}$. Very
approximately, after smoothing the mass retained in the initial area
will be of order of the fraction of the initial area to the one after
smoothing, $r^2 / (r + r_{\rm k})^2$.

After rearrangement (and using $r_{\rm k} \ll r$), we find that the
smoothing-induced mass bias should obey a relation of the form
\begin{equation}
\frac{\Delta m}{m} \approx  0.2 ~ {\rm to} ~ 0.3 \cdot
\left[ 1 -
\frac{1}{(1 + r_{\rm k} / r)^2} \right] \, .
\label{eq:uncertainty-resolution}
\end{equation}
To allow for more realistic source geometries, the numerical constant
must be derived from experiments with actual data. We use the
smoothing experiments shown in Fig.\
\ref{fig:comparison-noise-resolution}(b) for this purpose. With these
parameters, Eq.\ (\ref{eq:uncertainty-resolution}) describes the
relative mass bias for all structure branches in our example map.

In our study, we only consider regions with radii larger than twice
the effective beam radius, $r > r_{\rm beam}$. Substitution of this
(with $r_{\rm beam, i} = 0$, since the true distribution on the sky
has infinite resolution) in Eq.\
(\ref{eq:uncertainty-resolution}) shows that the mass of regions
with a radius similar to the beam diameter may be underestimated by
not more that 20\%, and by less for larger fragments.

\subsection{Combining Dust Emission and Extinction
  Data\label{sec:combine-masses}}
Our Perseus extinction map has a spatial resolution of $5 \arcmin$
($0.4 ~ \rm pc$). It cannot be used to characterize much smaller
fragments, such as dense cores of $\lesssim 0.1 ~ \rm pc$ size. This
limitation can, however, be overcome when including mass-size data
from additional data sets.

\subsubsection{Basic Concept}
For Perseus, \citet{enoch2006:perseus} present a Bolocam map of dust
emission in Perseus at $1.1 ~ \rm mm$ wave length. The beam width at
half sensitivity is $31 \arcsec$, corresponding to $0.04 ~ \rm pc$. At
these wave lengths, the continuum emission of dust at temperature
$T_{\rm d}$ is optically thin, and so the observed dust continuum
emission intensity, $I_{\nu}$, is a direct measure of the column
density along a given line of sight,
\begin{equation}
I_{\nu} = \mu_{\rm H_2} \, m_{\rm H} \, \kappa_{\nu} \,
N_{\rm H_2} \, B_{\nu}(T_{\rm d})
\label{eq:coldens-dust-emission}
\end{equation}
(see \citealt{kauffmann2008:mambo-spitzer} for full details), where
$B_{\nu}$ is the Planck function, and the dust opacity, $\kappa_{\nu}$,
is evaluated per total gas mass. Column density maps from dust
emission can then be contoured and analyzed as described before (Sec.\
\ref{sec:method}) to derive mass-size data for the fragments in the
dust emission map.

\begin{figure*}
\includegraphics[scale=0.55,bb=20 10 361 386,clip]{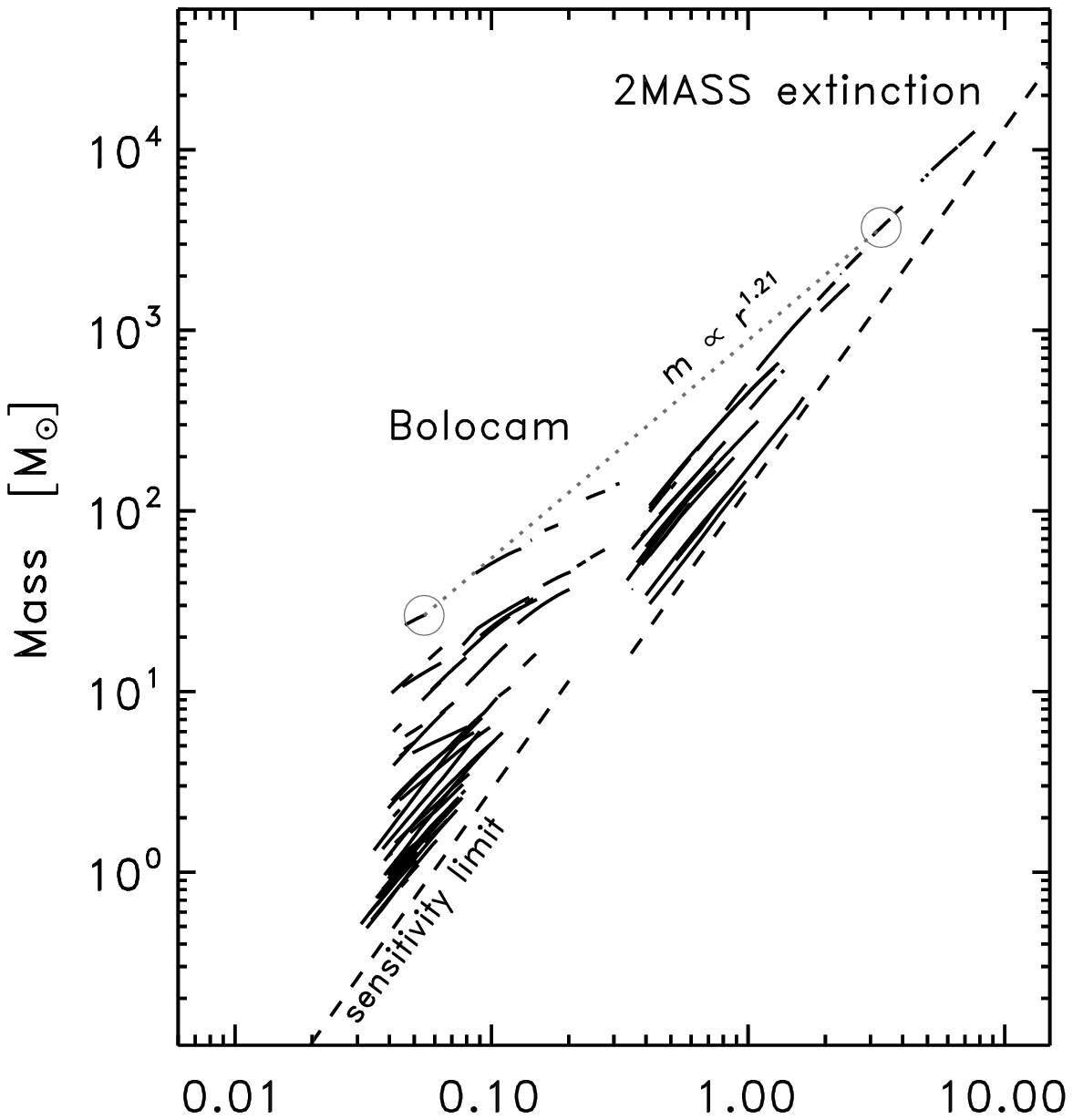}
\includegraphics[scale=0.55,bb=76 10 361 386,clip]{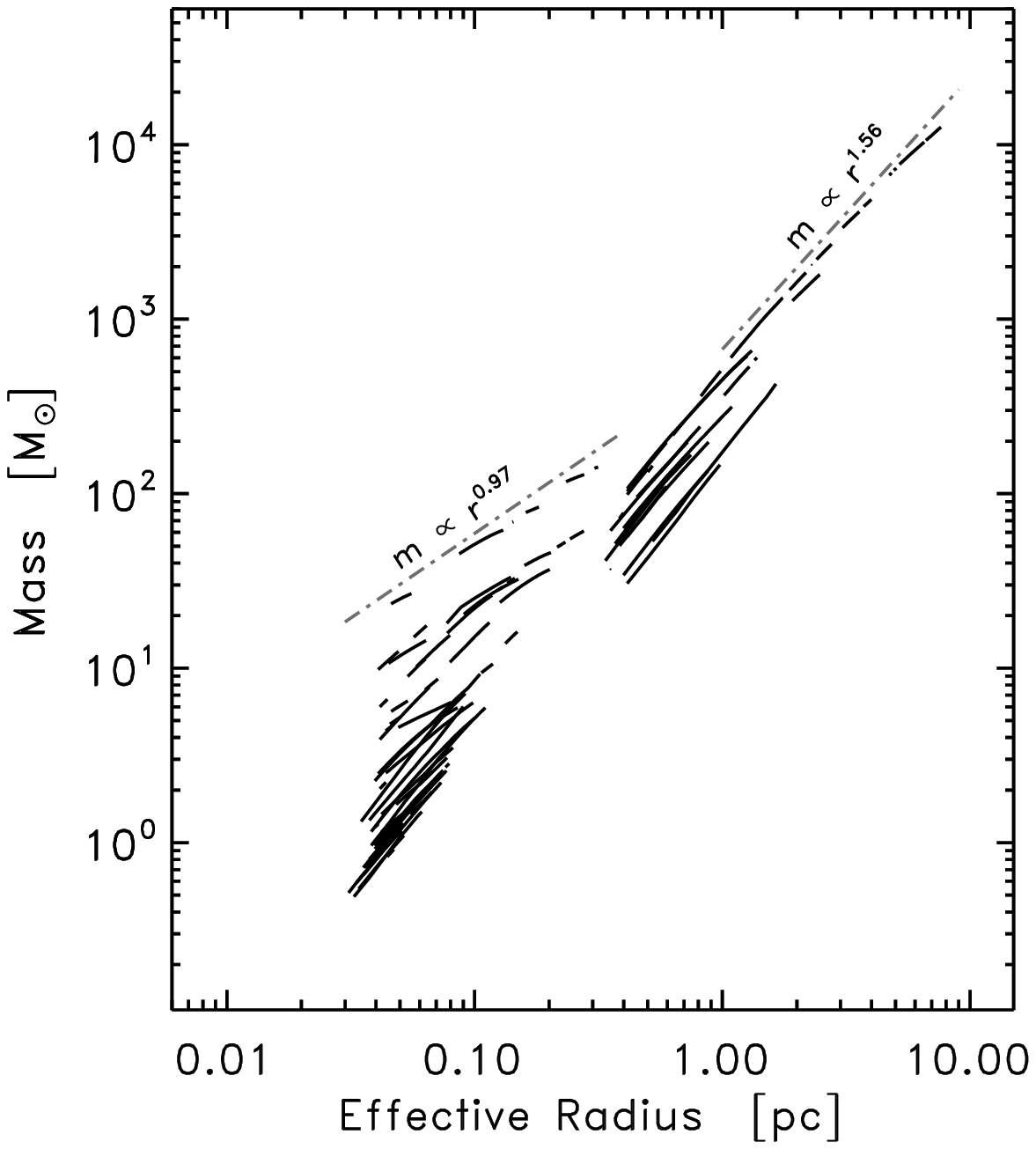}
\includegraphics[scale=0.55,bb=76 10 361 386,clip]{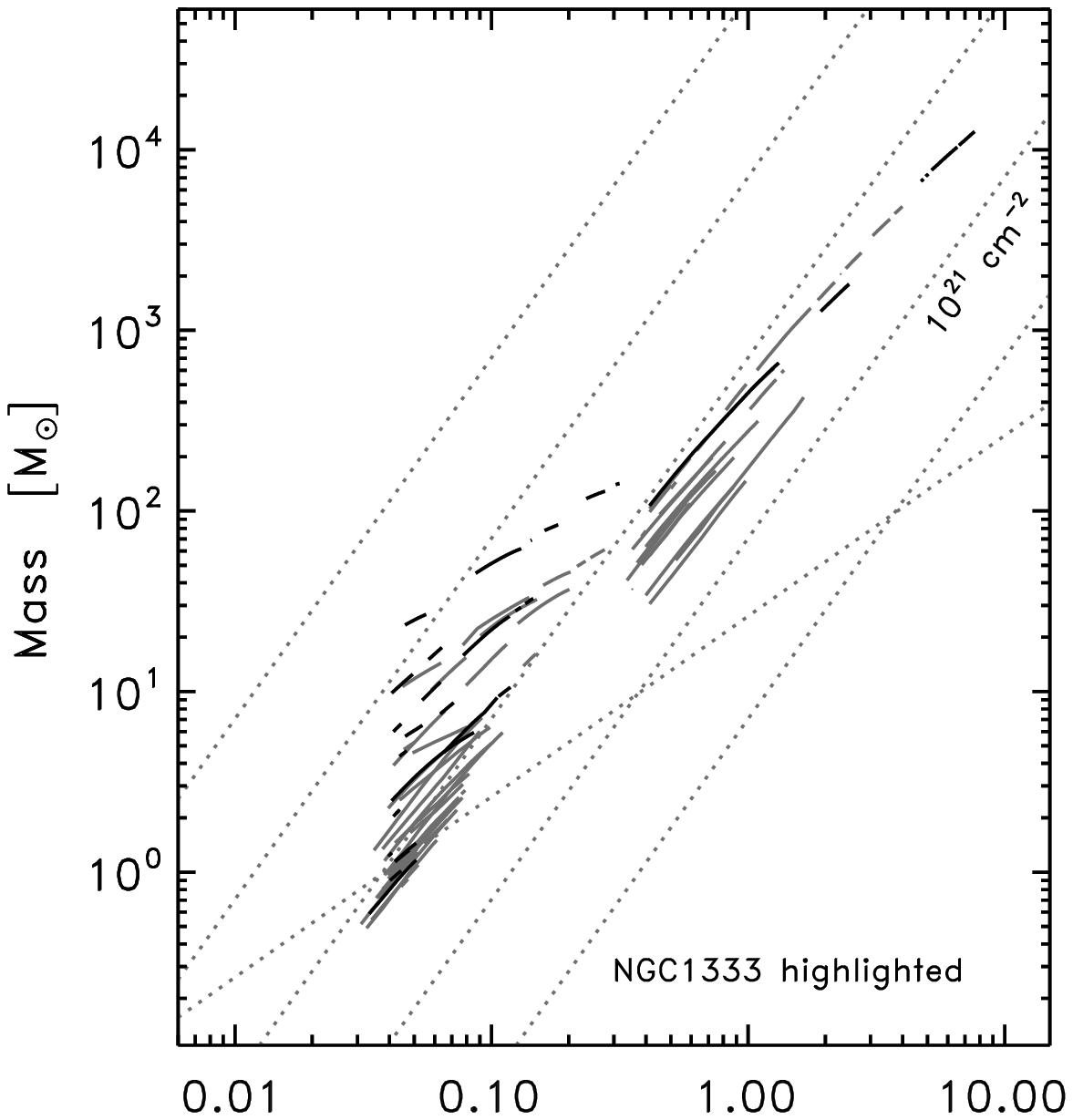}
\caption{Mass-size data for Perseus (\emph{solid lines}). The data are
  from dust extinction maps probing large spatial scales, and Bolocam
  dust emission observations sensitive to small ones. The \emph{left
    panel} highlights the nature of the data; \emph{dashed lines} give
  the instrument-dependent sensitivity limits. \emph{Circles and the
    dotted line} indicate calculation of the global mass-size slope
  (Sec.\ \ref{sec:slope_global-perseus}). The \emph{middle panel}
  highlights slopes derived by matching power-laws to parts of the
  data. Tangential slopes (Sec.\ \ref{sec:slopes-tangential}) are
  derived in the same fashion, but for infinitesimal radius
  ranges. The \emph{right panel} presents Perseus in the context of
  the reference mass-size relations from Sec.\
  \ref{sec:reference-relations}. Here, \emph{black solid lines}
  highlight Bolocam-detected fragments in NGC1333, as well as the
  extinction-probed fragments containing this cluster. Other data are
  drawn using \emph{grey solid lines}.\label{fig:emission-extinction}}
\end{figure*}

Figure \ref{fig:emission-extinction} combines the Bolocam-derived
mass-size measurements with the extinction-based ones. Spatially, they
are separated by a gap, since Bolocam ($\lesssim 2 \arcmin$) and 2MASS
extinction maps ($\gtrsim 5 \arcmin$) probe different spatial
scales. The extinction-identified fragments do, however, contain the
Bolocam-derived ones. For example, shading in Fig.\
\ref{fig:emission-extinction} highlights Bolocam-detected fragments
contained in the extinction peak harboring NGC1333.

\subsubsection{Relative Calibration of Dust Opacities}
\citet{bianchi2003:dust-emissivity} studied B68 to examine the
relation of mass estimates from dust extinction and emission. They
find relative differences by factors of $1 ~ {\rm to} ~ \sim 2.35$,
when adopting opacities for \citet{ossenkopf1994:opacities} dust
grains with thin ice mantles that coagulate for $10^5 ~ \rm yr$ at
$10^6 ~ \rm cm^{-3}$ density (see
\citealt{kauffmann2008:mambo-spitzer}, Table A.1, for numerical
values). This is consistent with previous studies
of this subject, including sources as diverse as diffuse clouds and
galaxies (see references in \citealt{bianchi2003:dust-emissivity}). If
one fixes the opacity used for extinction observations, it thus
appears that the the \citet{ossenkopf1994:opacities} model opacities
near $1 ~ \rm mm$ wave length are too large by an average factor
$1.5 \approx 2.35^{1/2}$.

In our emission-based column density estimates, we do therefore
basically adopt the aforementioned \citet{ossenkopf1994:opacities}
opacities for dust emission observations near $1 ~ \rm mm$ wave
length, but scale these down by a further factor 1.5 to bring mass
estimates from extinction into harmony with those from dust
emission. As seen in Fig.\ \ref{fig:emission-extinction}, this
procedure yields reasonable results, since the dust emission and
extinction observations for NGC1333 match within less than a factor 2
(i.e., when comparing the masses of the most massive Bolocam-detected
fragment to the one of the least massive 2MASS-identified
fragment). Incorrect assumptions about dust temperatures may cause
most of this offset (see below).

To some extent, correction factors between masses from dust emission
and extinction are also influenced by spatial filtering affecting
bolometer observations (Sec.\
\ref{sec:mass-uncertainties_bolo-filtering}). Future studies need to
address this problem in more detail. Still, the aforementioned scaling
factor aligns dust emission and extinction studies, which is the only
aspect needed in our present study.

\subsubsection{Dust Temperatures}
For Perseus, \citet{rosolowsky2007:perseus_nh3} estimate gas
temperatures between $9 ~ \rm K$ and $18 ~ \rm K$.  Assuming that dust
and gas temperatures are similar, we do therefore adopt a temperature
of $(12.5 \pm 2.5) ~ \rm K$ in our mass estimates.  This temperature
uncertainty results in a relative mass uncertainty of $\sim 20\%$ for
Perseus. A few sources may have dust temperatures, and mass biases,
outside of this range.

\subsubsection{Impact of Spatial
  Filtering\label{sec:mass-uncertainties_bolo-filtering}}
Like all other upcoming, present, and past ground-based
bolometer-derived dust emission maps
\citep{kauffmann2008:mambo-spitzer}, the Bolocam maps of Perseus are
not sensitive to structure larger than some instrument-dependent
spatial scale. In its impact, this problem is similar to the spatial
filtering in interferometric imaging. For Bolocam,
\citet{enoch2006:perseus} show the filtering scale to be of order
$1 \arcmin$ to $2 \arcmin$ radius.

Quantitatively, this removal of large-scale structure has an influence
opposite to the impact of smoothing (Sec.\
\ref{sec:uncertainty-mass}): here, the relative mass-loss increases
with spatial scale. This bias has to be considered carefully when
using emission-based mass-size measurements for analysis. Obviously,
the true mass will be larger than the observed value. Filtering in
bolometer maps is unfortunately too complex to be characterized in a
compact fashion; see \citet{kauffmann2008:mambo-spitzer} for a few
rules of thumb. For the particular case of the Bolocam maps of
Perseus, \citet{enoch2006:perseus} report losses $\le 10\%$ for
radii $\le 1 \arcmin$, but do not sufficiently explore larger
objects.

\subsection{Global Trends in Maximum Mass for given
  Size\label{sec:slope_global-perseus}}
The mass-size tendencies seen, e.g., in Fig.\
\ref{fig:emission-extinction} suggest to describe these trends with
power laws of the form
\begin{equation}
m(r) = m_0 \, (r / r_0)^b \, .
\label{eq:power-law}
\end{equation}
In this, $b$ is the slope of the relation, and $m_0$ is the
intercept. As shown in Fig.\ \ref{fig:emission-extinction}, such laws
trace, e.g., the mass-size relation of the most massive fragments in
the $0.04 \le r / {\rm pc} \le 0.3$ and $1 \le r / {\rm pc} \le 8$
radius intervals with maximum deviations of $\pm 6\%$.

Beyond such detailed descriptions of individual features, power-laws
can be used to capture more global aspects of a cloud's
structure. Consider, for example, the relation between the maximum
fragment masses, $m_{\rm max}$, observed at radii of
$r_{\rm sm} = 0.05 ~ \rm pc$ and $r_{\rm lg} = 3.0 ~ \rm pc$. (These
radii are chosen to permit comparison with other clouds, as becomes
more obvious below.) Based on these, we can define a global slope,
\begin{equation}
b_{\rm glob} = \frac{
  \ln[m_{\rm max}(r_{\rm lg}) / m_{\rm max}(r_{\rm sm})]
}{
  \ln[r_{\rm lg} / r_{\rm sm}]
} \, .
\label{eq:slope-global}
\end{equation}
As illustrated in Fig.\ \ref{fig:emission-extinction}, this slope is
defined such that Eq.\ (\ref{eq:power-law}) connects the mass
and size measurements for $b = b_{\rm glob}$. In Perseus,
$b_{\rm glob} = 1.21 \pm 0.14$, where we evaluate the
uncertainty very conservatively by scaling the emission-based mass in
Eq.\ (\ref{eq:slope-global}) up and down by a factor 2.

In Perseus, at the chosen scale of $r = 0.05 ~ \rm pc$ (i.e.,
$40\arcsec$), the Bolocam dust emission map is not significantly
affected by spatial filtering. Similarly, at $r = 3 ~ \rm pc$
($40\arcmin$), the extinction map is not biased by presence of stellar
clusters. The measurements of $b_{\rm glob}$ are thus immune to these
influences.

\subsection{Local Trends in Mass\label{sec:slopes-tangential}}
We now turn to slopes of infinitesimal tangents. As we explain below,
the Bolocam data are not suited for this analysis, since they suffer
from too strong spatial filtering.

\subsubsection{Method and Uncertainties}
\begin{figure}
\begin{tabular}{cc}
\includegraphics[width=0.92\linewidth,bb=20 55 360 200,clip]{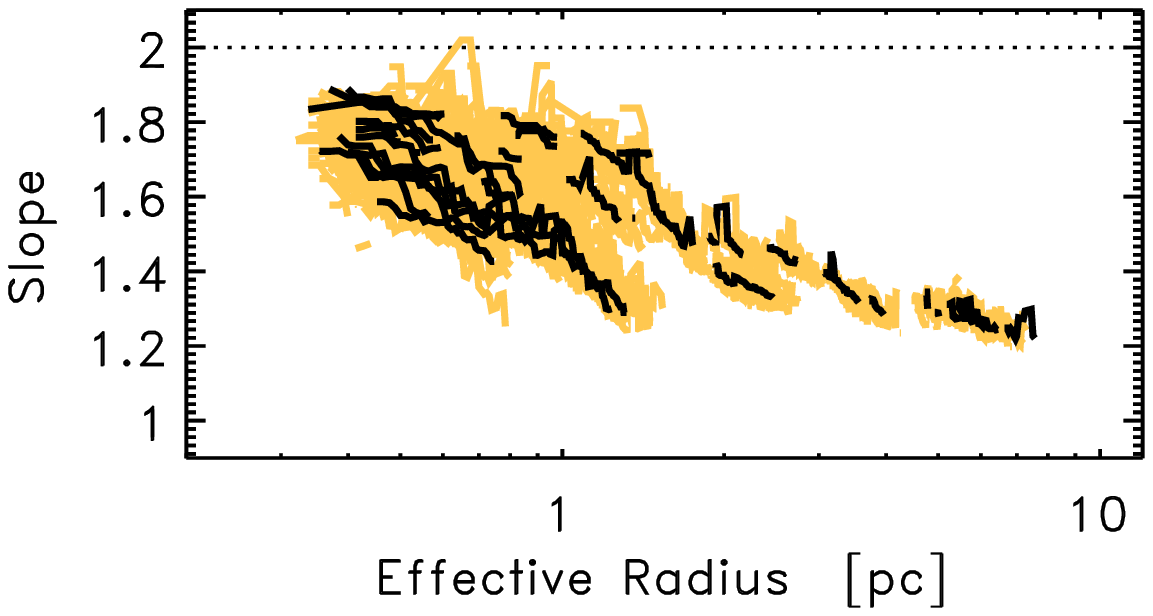} &
\begin{sideways}
\textsf{\hspace{0.4cm}a) plain differences}
\end{sideways}\\
\includegraphics[width=0.92\linewidth,bb=20 55 360 188,clip]{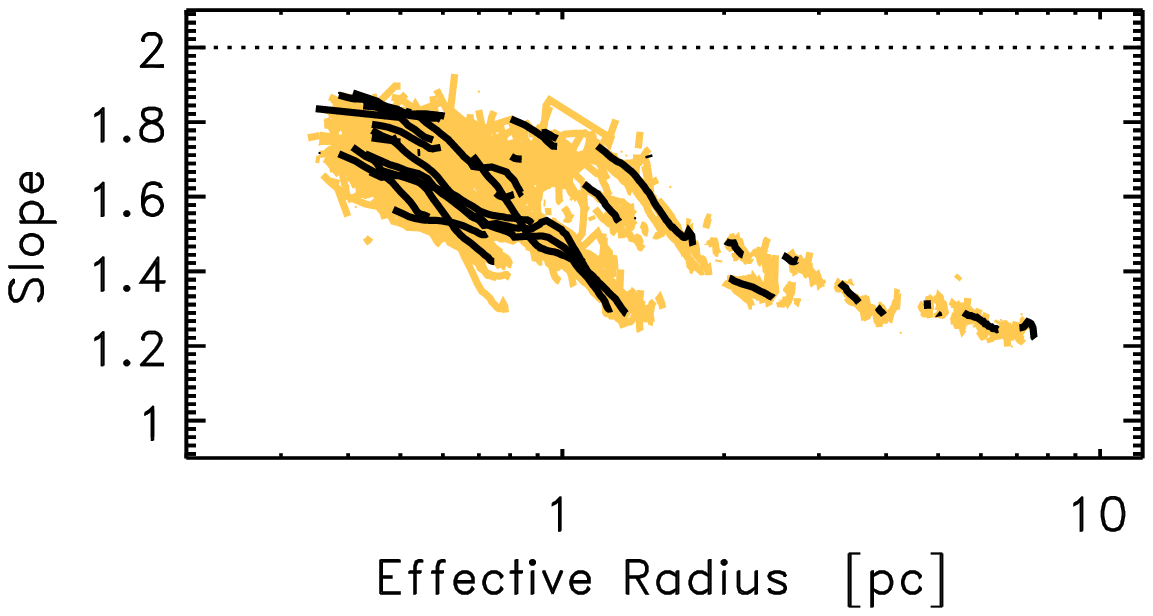} &
\begin{sideways}
\textsf{\hspace{0.7cm}b) smoothed}
\end{sideways}\\
\includegraphics[width=0.92\linewidth,bb=20 10 360 188,clip]{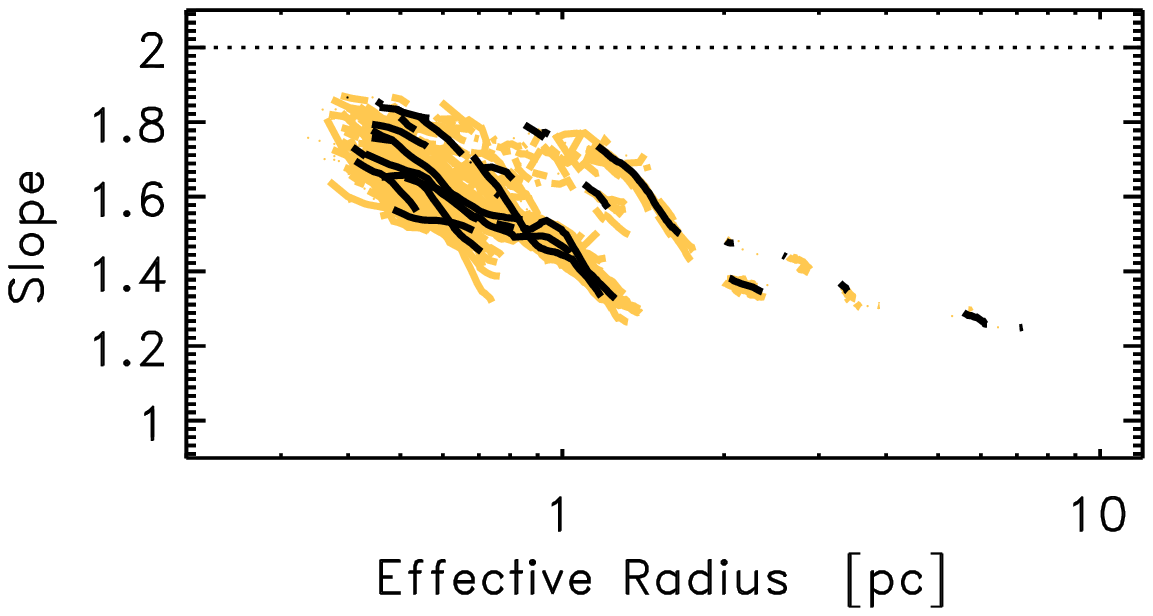} &
\begin{sideways}
\textsf{\hspace{1.cm}c) smoothed and filtered}
\end{sideways}
\end{tabular}
\caption{Calculation scheme for slopes (\emph{top to bottom}) and
  impact of noise (\emph{black vs.\ yellow lines}). We start with slopes
  directly calculated from mass and size differences for
  successive contours (panel a). These measurements are subsequently
  smoothed (panel b), and data are removed where the smoothing kernel
  is not filled (panel c). The \emph{black lines} hold for the
  observed Perseus extinction data. As in  Fig.\
  \ref{fig:comparison-noise-resolution}, \emph{yellow and black lines}
  indicate results from maps with and without artificial noise,
  respectively. The \emph{dotted line} indicates the upper limit on
  slopes inherent to our method, $b < 2$ (Eq.\
  \ref{eq:mr_column-density}).\label{fig:comp-slopes-smoothing}}
\end{figure}

In some cases, one may wish to use Eq.\ (\ref{eq:power-law}) to
describe tangents to the data on spatial scales smaller than the one
for which we calculate global slopes. This is demonstrated in Fig.\
\ref{fig:emission-extinction} using tangents to the most massive
fragments seen at given radius. In particular, it can be desirable to
fit infinitesimal tangents to mass-size trends. For these, the slope
reads
\begin{equation}
b(r) =
\left.
\frac{{\rm d} \, \ln(m[r'])}{{\rm d} \, \ln(r')}
\right|_{r' = r} \, .
\label{eq:slopes_tangential}
\end{equation}
Figure \ref{fig:comp-slopes-smoothing}(a) shows slopes derived from
mass and size differences between consecutive contours. As seen in the
figure, these data are rather noisy. We do therefore smooth the
measurements. To do this, at radius $r$, we replace slope and radius
by their respective unweighted arithmetic means, as derived within a
smoothing kernel of width $[r, 1.15 \cdot r]$ (for a
given cloud fragment, not permitting mergers). This yields the data
shown in Fig.\ \ref{fig:comp-slopes-smoothing}(b). Sometimes, however,
the smoothing kernel is not filled well, and the data are still
noisy. Thus, we do finally remove all data where the kernel is not
filled to at least $2/3$. Figure \ref{fig:comp-slopes-smoothing}(c)
shows this final result.

The impact of noise can be estimated by propagating the mass
uncertainties due to noise (Eq.\ \ref{eq:uncertainty-noise}) within
the slope calculations. This yields
\begin{equation}
\frac{\sigma(b)}{b} \le 20 \,
\frac{\sigma(m_i)}{m} \, \frac{r}{r_{\rm beam}} \, .
\label{eq:uncertainty-smooting-filtering_slopes}
\end{equation}
Because of the aforementioned smoothing, the numerical constant must
be calibrate with our noise experiments (as done for Eq.\
[\ref{eq:uncertainty-noise}]). To obtain an estimate of the expected
uncertainties, we can repeat the mass and size substitutions done in
the discussion of Eq.\ (\ref{eq:uncertainty-noise}). This gives
maximum uncertainties $\sim 60\%$. Since
$\sigma(b) / b \propto \langle N_{\rm H_2} \rangle^{-1} \, r^{-1}$
(see discussion of Eq.\ [\ref{eq:uncertainty-noise}]), the
uncertainties are small for larger regions well above the detection
threshold. In practice, uncertainties $< 10\%$ are a reasonable
estimate for well-detected regions warranting detailed study.

The slope difference due to smoothing is given by the first derivative
of the mass bias due to smoothing (Eq.\
\ref{eq:uncertainty-resolution}) with respect to the radius. Including
the usual calibration of numerical constants, we constrain the
absolute smoothing-induced error to
\begin{equation}
\Delta b \le 0.4 ~ {\rm to} ~ 0.6 \,
\frac{r_{\rm k} / r}{(1 + r_{\rm k} / r)^3} \, .
\label{eq:uncertainty-resolution_slope}
\end{equation}
In a few cases, however, the error can be larger by a factor 2.  In
this paper, we reject regions with a diameter smaller than twice the
beam diameter. Substitution of this limit (i.e.,
$r_{\rm k} / r < 1/2$) into Eq.\
(\ref{eq:uncertainty-resolution_slope}) implies that slopes are
overestimated by a number smaller than 0.1 due to resolution.

\begin{figure}
\includegraphics[width=\linewidth,bb=20 10 360 200,clip]{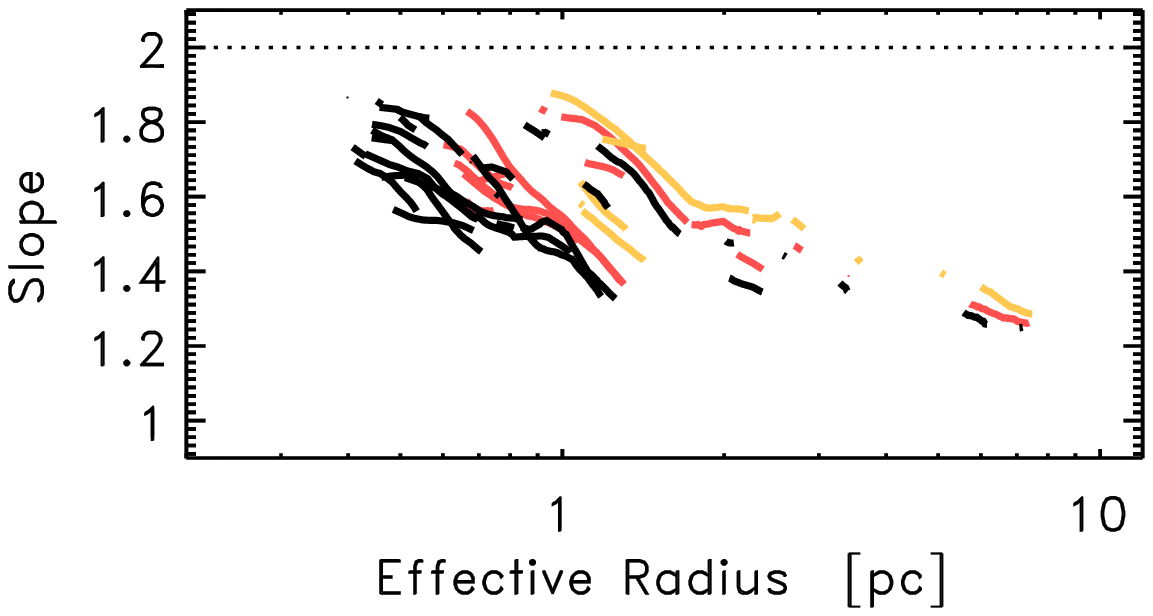}
\caption{Impact of limited spatial resolution on slope
  measurements. See Figs.\ \ref{fig:comparison-noise-resolution}(b)
  and \ref{fig:comp-slopes-smoothing} for explanations of
  mark-up.\label{fig:comp-slopes-resolution_noise-multi}}
\end{figure}

As mentioned in Sec.\ \ref{sec:mass-uncertainties_bolo-filtering},
bolometer maps suffer from spatial filtering. This has an impact
opposite to the influence of smoothing, and artificially shallow
slopes are measured from such maps. Since the filtering is very strong
in Bolocam maps, we do not use these for the derivation of tangential
slopes.

\subsubsection{Results for Perseus\label{sec:slopes-infinitesimal-perseus}}
Figures \ref{fig:comp-slopes-smoothing}(c) and
\ref{fig:comp-slopes-resolution_noise-multi} show tangential slopes
for Perseus. The tangential slopes in the $1 \le r / {\rm pc} \le 8$
radius interval are in the range $b = 1.2 ~ {\rm to} ~ 1.7$.
At a given radius, the slopes for different fragments do often differ
by more than their uncertainty. Also, in a given fragment the slope
can change significantly with respect to radius. This means that it is
not possible to describe an entire cloud by a single tangential
slope. The observed tangential slopes bear no obvious relation to the
global slope $\sim 1.56$ derived for the same radius range (Fig.\
\ref{fig:emission-extinction}).

A slope of $\ge 2$ is actually not possible for a mass-size relation
as defined by us, since this would require that the column density
increases with radius (Eq.\ \ref{eq:mr_column-density}). Since our
search algorithm proceeds by decreasing the column density threshold,
such fragments are not identified by our identification scheme.  A
slope $b < 0$ would mean that the mass decreases with increasing
radius. This is impossible in all but the most jolly insane models of
cloud structure.

\section{Summary \& Outlook}
\subsection{Utility of and Outlook for Mass-Size
  Studies\label{sec:purpose}}
As we have shown above, our map characterization scheme yields
reliable measurements of mass and size. From these, mass-size slopes
and intercepts can be derived. Below, we describe how these data
contribute to critical fields of star formation research.\medskip

First of all, this approach permits a \emph{continuous}
characterization of cloud structure across a large range of spatial
scales. This is just a desirable feature of any data analysis method,
independent of the exact nature of the later analysis. The need for
such a procedure led \citet{rosolowsky2008:dendrograms} to the
development of the ``dendrogram technique''.

In star formation research, the basic mass-size measurements permit to
compare fragment masses at a given spatial scale. Consider the
classical case in order to see the advantage: usually, ``cores'' and
``clumps'' extracted from maps differ in their size. In this case, it
is not clear what differences in masses mean, even if just a single
cloud is considered.

Spatially continuous cloud characterizations become particularly
useful when comparing observations for different molecular
clouds. Usually, every cloud is studied at a different physical
resolution (i.e., pc per pixel). In the classical case, mass
measurements will thus usually refer to vastly different spatial
scales. With our approach, however, all scales are probed, and
different clouds can be compared at the same physical scale. This is
extensively employed in part II, where we study a sample of
clouds.\medskip

The general utility of measurements of mass and size is known since
long. For example, one can compare the actual to virial masses or,
more generally, masses predicted by theoretical cloud models. Equation
(\ref{eq:mass-size-sis}), for example, relates model fragment masses
and gas temperatures. We shall not discuss such considerations here in
detail.

A property \emph{uniquely} constrained by our method are mass-size
slopes; these can \emph{only} be measured via a scale-independent
method. In particular, this gives access to the density structure of
molecular clouds. For simple models of cloud structure, the mass-size
slope is, e.g., directly related to the slope of the density profile
(Eq.\ \ref{eq:slope-vs-density}). Such work on large-scale structure
in molecular clouds is urgently needed, since cloud density profiles
are presently not known on scales $\gtrsim 0.1 ~ \rm pc$. Part II
presents slope measurements for various molecular clouds, as well as
several model mass-size relations.

\subsection{Summary\label{sec:summary}}
This work studies the internal structure of molecular clouds by
breaking individual cloud complexes up into several nested
fragments. For these, we derive masses and sizes, as e.g.\ outlined in
Fig.\ \ref{fig:processing-scheme}. Effectively, we perform a
``dendrogram analysis'' of a two-dimensional map, as introduced by
\citet{rosolowsky2008:dendrograms}.

The present paper establishes the method via a detailed analysis of
the Perseus Molecular Cloud. Other solar neighborhood molecular clouds
($\lesssim 500 ~ \rm pc$; the Pipe Nebula, Taurus, Ophiuchus, and
Orion) are discussed in the next paper of this series (part II).

Power-laws of the form $ m(r) = m_0 \, (r / {\rm pc})^b$,
with slope $b$ and intercept $m_0$, prove useful to quantify
the relations between mass, $m$, and size, $r$ (i.e., the
effective radius). Sections \ref{sec:slope_global-perseus} and
\ref{sec:slopes-tangential} discuss two different approaches to define
and measure the slope. We use \emph{global slopes} to measure the
relation between structure at small and large scales. This is done by
connecting the mass-size measurements of the most massive fragments at
small ($0.05 ~ \rm pc$) and large radius ($3.0 ~ \rm pc$) by a
power-law (Eq.\ \ref{eq:slope-global}). \emph{Tangential slopes}, on
the other hand, are calculated infinitesimally at a given spatial
scale (Eq.\ \ref{eq:slopes_tangential}). The uncertainties in these
properties are examined in Secs.\ \ref{sec:uncertainty-mass} and
\ref{sec:combine-masses} (for mass and intercept), respectively Secs.\
\ref{sec:slope_global-perseus} and \ref{sec:slopes-tangential} (for
slopes).

We conclude that our mass, slope, and intercept measurements provide a
reliable method to characterize cloud structure. Our approach enables
a continuous and reliable characterization of cloud structure in the
$0.05 \lesssim r / {\rm pc} \lesssim 10$ spatial range. This is not
possible using previous methods, since these are usually biased
towards a particular spatial scale (see, e.g., the CLUMPFIND analysis
in Fig.\ \ref{fig:comp-clumpfind}). Such comprehensive pictures of
star-forming regions can be used to develop a more complete
theoretical understanding of global cloud structure (Sec.\
\ref{sec:purpose}).

A first observational and theoretical exploitation of this method is
presented in part II of this series. We characterize, for example, the
typical parameter space for solar neighborhood molecular clouds not
forming massive stars. Based on this, we chart a potential mass-size
threshold for the formation of massive stars. Mass-size slopes are
used to constrain large-scale density gradients within molecular
clouds.

\acknowledgements{We are grateful to Nicolas Peretto, who served as a
  considerate and knowledgeable referee who helped to significantly
  improve the quality of the paper. This project would not have been
  possible without help from Erik Rosolowsky. His dendrogram analysis
  code \citep{rosolowsky2008:dendrograms} was instrumental for our
  analysis. We thank Jaime Pineda for his help with the CLUMPFIND
  experiments presented in Fig.\ \ref{fig:comp-clumpfind}.
  \citet{enoch2006:perseus} contributed maps to the present study. We
  are grateful for their help. This work was in part made possible
  through Harvard Interfaculty Initiative funding to the Harvard
  Initiative in Innovative Computing (IIC).}

\bibliographystyle{apj}
\bibliography{mendeley/bib_astro-Mendeley}

\end{document}